\documentclass[prl,twocolumn,floatfix,showpacs,amsmath,amssymb,superscriptaddress]{revtex4}
\usepackage{graphicx}
\begin{document}


\title{Efficient Site-specific Low-energy Electron Production 
via Interatomic Coulombic Decay Following Resonant Auger Decay}
\author{M.~Kimura}
\affiliation{Institute of Multidisciplinary Research for Advanced
Materials, Tohoku University, Sendai 980-8577, Japan}
\author{H.~Fukuzawa}
\affiliation{Institute of Multidisciplinary Research for Advanced
Materials, Tohoku University, Sendai 980-8577, Japan}
\author{K.~Sakai}
\affiliation{Institute of Multidisciplinary Research for Advanced
Materials, Tohoku University, Sendai 980-8577, Japan}
\author{S.~Mondal}
\affiliation{Institute of Multidisciplinary Research for Advanced
Materials, Tohoku University, Sendai 980-8577, Japan}
\author{E.~Kukk}
\affiliation{Department of Physics, University of Turku, FI-20014 Turku, Finland}
\author{Y.~Kono}
\affiliation{Department of Chemistry, Ehime University, Matsuyama 790-8577, Japan}
\author{S.~Nagaoka}
\affiliation{Department of Chemistry, Ehime University, Matsuyama 790-8577, Japan}
\author{Y.~Tamenori}
\affiliation{Japan Synchrotron Radiation Research Institute, Sayo,
Hyogo 679-5198, Japan}
\author{N.~Saito}
\affiliation{National Institute of Advanced Industrial Science and
Technology, NMIJ, Tsukuba 305-8568, Japan}
\author{K.~Ueda}\email[]{ueda@tagen.tohoku.ac.jp} \affiliation{Institute
of Multidisciplinary Research for Advanced Materials, Tohoku
University, Sendai 980-8577, Japan}
\date{\today}


\begin{abstract}
We identified interatomic Coulombic decay (ICD) channels in argon 
dimers after spectator-type resonant Auger decay $2p^{-1}~3d \to 
3p^{-2}3d, 4d$ in one of the atoms, using momentum resolved 
electron-ion-ion coincidence. 
The results illustrate that the resonant core excitation is 
a very efficient way of producing slow electrons at a specific site, 
which may cause localized radiation damage.
We find also that ICD rate for $3p^{-2}4d$ is significantly 
lower than that for $3p^{-2}3d$. 
\end{abstract}

 \maketitle



Though inner-valence vacancy states in molecules are usually 
not subject to autoionization, such states 
may be subject to autoionization if the inner-valence ionized
molecule is close to other molecules. 
This new type of autoionization was noted by Cederbaum 
{\em et al.}~\cite{Cederbaum97} and called intermolecular or 
interatomic Coulombic decay (ICD).  
Since then, many theoretical and experimental studies have 
been reported in many different systems. 
See, e.g., recent review articles~\cite{Averbukh,Hergenhahn}.  
It is now well-known that ICD appears everywhere and transfers 
the energy and the charge from the excited species to the 
environment surrounding it. 
Also it is known that there are many variants of ICD, 
such as ICD from satellite 
states~\cite{Jahnke07,Lablanquie07,Havermeier10,Sisourat10},
ICD after Auger 
decay~\cite{Santra03,Morishita06,Morishita08,Kreidi08a,Yamazaki08,Kreidi08b,Kreidi09}, 
resonant ICD~\cite{Barth05,Aoto06,Gokhberg06} 
where inner-valence excited states decay via ICD-like spectator decay,
and 3-electron ICD~\cite{Averbukh09,Ouchi11} 
where two inner-valence holes in a species are filled by ICD 
in which three electrons are involved.
Another exotic variant, called electron-transfer-mediated decay 
(ETMD), was also predicted~\cite{Zobeley01} 
and observed~\cite{Sakai11,Forstel11}. 
 
Relevance of ICD processes with radiation damage is also 
noted~\cite{Morishita06,Jahnke10,Mucke10,Vendrell10,Stoychev11,Kim11}.
Noting that ICD produces low energy electrons locally at the site 
where an excitation takes place, one mayaW think one step further: 
one may consider the relevance of  the production of low energy 
electron via ICD with radiation therapy
that requires the localized radiation damage. 
Up to now, however, ICD has been studied with ionizing radiation 
that does not have a selectivity of the site or the states. 
Namely, in order to observe ICD, one usually ionizes the inner-valence 
orbital or populate the satellite state by the monochromatic 
radiation~\cite{Jahnke07,Jahnke10,Mucke10}.  
This radiation, however, not only populates the ICD initial state 
of the target atom or molecule at a specific site 
but also ionizes inner- and outer-valence orbitals 
of any atoms and molecules.  
Using core ionization by the radiation with photon energy 
above a core ionization potential of the specific atom 
may improve, to a certain amount, 
the selectivity of the atomic photoabsorption.  
ICD takes place from the Auger final states that 
have the inner-valence hole~\cite{Santra03}.
The branching ratio to these states are, however, 
only of the order of $10$ \%~\cite{Morishita06,Kreidi08b,Ouchi11}. 
In turn, a well-established method of site-specific excitation is
resonant core excitation; using the chemical shift, one can select not only 
a specific kind of atoms but also a specific site of the same kind of atoms 
located at different sites.  
Very recently, Gokhberg {\it et al.} suggested this pathway: 
the merits of this ICD mode are the high site-selectivity and 
the tunability of the ICD-electron energy using different resonant core 
excitations~\cite{Gokhberg}.

\begin{figure}[tpb]
\begin{center}
\includegraphics[width=8cm]{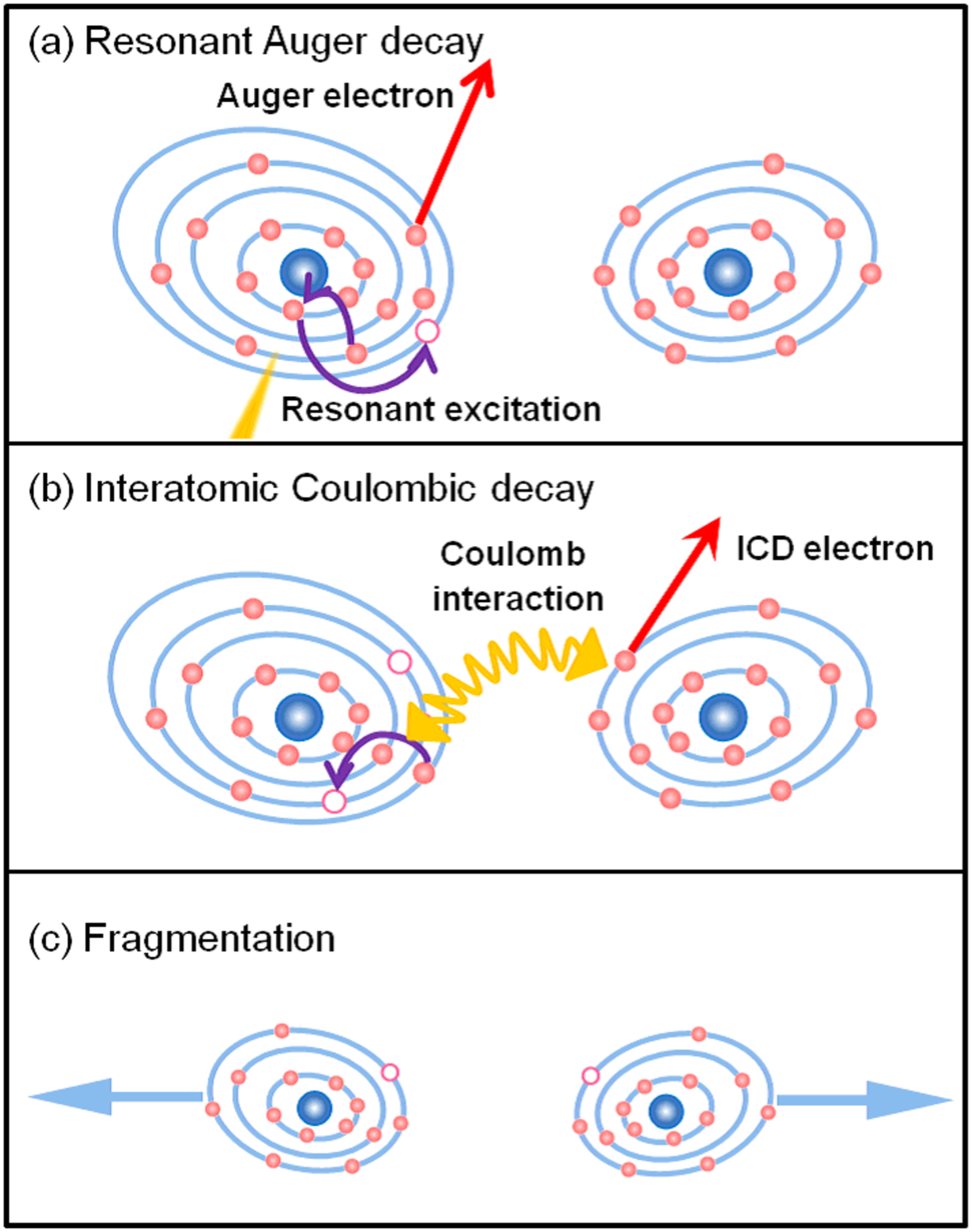}
\end{center}
\caption{(Color on line.) Sequence of events observed in the Ar dimer.
(a)~Photoexcitation promotes a 2$p$ electron to an unoccupied 
$3d$ orbital in one of the atoms. The 2$p$ vacancy is filled by one 
of the 3$p$ electrons and another 3$p$ electron is emitted 
from the same atom, while the 3$d$ electron remains as a spectator 
(resonant Auger decay).
(b)~Interatomic Coulombic decay takes place, in which one of the 
two 3$p$ vacancies is filled by the 3$d$ electron from the same atom
and the excess energy is transferred to the neighboring atom which
in turn emits one of its 3$p$ electrons. (c)~Fragmentation due to
Coulomb explosion takes place. The Ar$^+(3p^{-1})$ and 
Ar$^+(3p^{-1})$ fragment
ions along with the ICD electron emitted in (b) are detected in
coincidence.} \label{fig1}
\end{figure}

In the present work, 
we demonstrate that ICD indeed takes place following resonant Auger decay 
efficiently, with a probability close to 100 \%. 
As a prototype sample, we have used argon dimers. 
The schematic sequence we have investigated is illustrated in Fig.~1.  
When a 2$p$ electron in one of the atoms is excited to a 3$d$ orbital,  
a spectator Auger decay takes place in the atom, populating mostly
the spectator Auger final state 3$p^{-2}3d$ and its shakeup state 
3$p^{-2}4d$ [Fig.~1(a)].  
Then ICD takes place [Fig.~1(b)] from these states, 
leading to fragmentation 
to Ar$^+(3p^{-1})$-Ar$^+(3p^{-1})$ [Fig.~1(c)].  
In the experiment, we record 3D momentum for each of two 
Ar$^+$ ions and the ICD electron in coincidence.

The experiment was carried out on the c branch of the beam 
line 27SU~\cite{Ohashi01a,Ohashi01b,Ueda03} at SPring-8.  
The storage ring was operated 
in several-bunches mode providing 53 single bunches (4/58 filling
bunches) separated by 82.6 ns. The monochromatic radiation 
was directed horizontally, with vertical linear polarization. The
argon dimers Ar$_2$ were produced by expanding argon gas at
a stagnation pressure of 0.3 MPa at room temperature 
through a pinhole of 30~$\mu$m diameter. 
Under these conditions the cluster
beam contains monomers, dimers, and a very small fraction of larger
clusters. The signals from dimers were selected by applying the
momentum conservation law for the ion pairs detected in coincidence 
(see below).  The cluster beam was directed vertically and crossed the
incident radiation at right angles.  The incident energy was tuned to the
excitation energy 2$p_{3/2} \to 3d$ at 246.94 eV~\cite{Kato07JES}, 
with the photon band width of 0.13 eV.

Our momentum-resolved electron-ion 
multicoincidence~\cite{Morishita06,Ueda08} is equivalent to cold-target
recoil-ion momentum spectroscopy or reaction microscope~\cite{Ullrich03} 
and is based on recording times of flight (TOFs) for electrons and ions 
with two position and time sensitive multihit-capable detectors 
(Roentdek HEX120 for electrons and HEX80 for ions).
Knowledge of position and arrival time on the particle detectors,
$(x,y,t)$, allows us to extract information about the 3D momentum
of each particle. The electron and ion spectrometers were placed
face to face. The spectrometer axis was horizontal and
perpendicular to both the incident radiation and the cluster beam.
Electric and magnetic fields were applied to the interaction
region, so that all the electrons with energy up to 22 eV and all
the fragment ions were guided to the electron and ion detectors,
respectively.  Detailed geometric descriptions and typical field
conditions of the spectrometers were given
elsewhere~\cite{Ueda08}. The TOFs of electrons and ions were
recorded with respect to the bunch marker of the light source
using multi-hit time-to-digital converters (Roentdek TDC8HP),
selecting by logic gating only electron signals synchronized with
the single bunches.

\begin{figure}[tpb]
\begin{center}
\includegraphics[width=8cm]{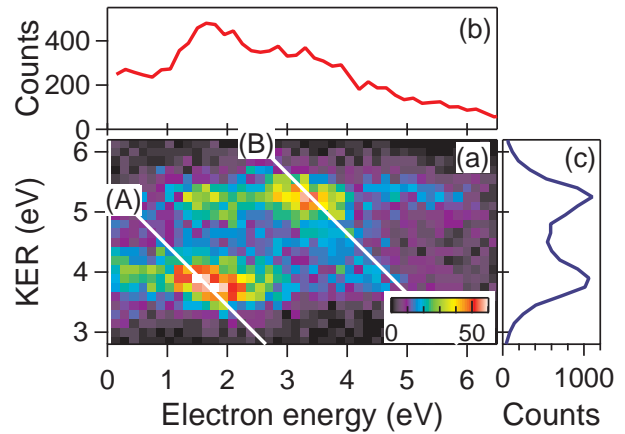}
\end{center}
\caption{(Color on line.) (a) Relationship between the electron energy and the
total kinetic energy release (KER) of the Ar$_2$ fragmentation. (b)
Electron energy distribution of the electron ejected from
Ar dimers. (c) The KER of the Ar$_2$ fragmentation.} \label{fig2}
\end{figure}

The coincidence measurement for one electron and two ions provides
the electron kinetic energy together with the kinetic energy release (KER)
between the two Ar$^+$ ions for each event.  
The relationship of the electron energy and 
the KER is shown in Fig.~2(a).  
There are two prominent islands; one at electron energy of 1.8 eV and 
KER = 3.8 eV and the other at electron energy of 3.3 eV and KER = 5.2 eV. 
We may notice also that there is one more faint island  
at electron energy of  1.8 eV and KER = 5.2 eV.  
Figure 2(b) shows the electron energy distribution recorded in 
coincidence with the fragmentation into Ar$^{+}$-Ar$^+$.  
These electrons correspond to the ICD electrons, as we will discuss 
below.  
Figure 2(c) shows the distribution of the KER. There are two peaks, 
one at 3.8 eV and the other at 5.2 eV.  If we assume the Coulomb 
repulsion between the two ions, 
then the corresponding internuclear distances are 3.8~\AA~and 2.8~\AA,
respectively. The bond-length of the 
neutral argon dimer is 3.76~\AA~\cite{Ogilvie92}. 
Thus the peak at 3.8 eV corresponds to the case where ICD 
takes place at the equilibrium internuclear distance of the neutral argon 
dimer.  The peak at 5.2 eV, on the 
other hand, corresponds to the case where ICD takes place at the
internuclear distance significantly shorter (by $\sim 1$~\AA) 
than the equilibrium bond length.

\begin{figure}[tpb]
\begin{center}
\includegraphics[width=8cm]{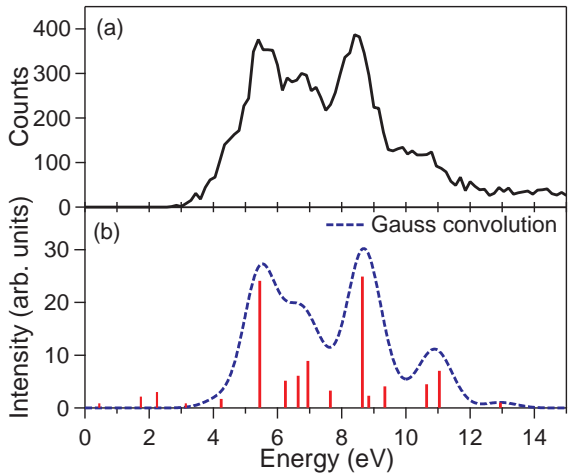}
\end{center}
\caption{(Color on line.)  
Energy distribution for the sum of the electron energy and the KER, 
as measured (a) and as estimated (b). The estimates given by 
the red bars in (b) are based on the resonant Auger transitions 
reported in~\cite{Meyer91}. 
See text and Table I for the details.  
The (blue) curve is a convolution of vertical lines by Gaussian 
profiles with 0.94 eV FWHM (i.e., the average experimental resolution) 
to compare with the experimental spectrum in (a). In the convolution 
the states below 4 eV given by thin red bars are excluded because 
the ICD channels are energetically closed.} \label{fig3}
\end{figure}

\begin{table}
\caption{Candidates of the ICD initial states, expected sum of 
electron kinetic energy and KER and relative intensities.  
The energies and the intensities are 
estimated from resonant Auger energies and intensities 
reported in \cite{Meyer91}. 
The ICD channels are energetically closed for the states below 4 eV.}
\begin{tabular}{rlcc}
\hline
  &  ICD initial state  & Energy & Relative \\
  &                     & (eV)   & intensity\\
\hline
3$p^4$($^3P$)3$d$ & $^4D$                & 0.44  & 0.86 \\
                  & $^4F$+$^2P$          & 1.74  &  2.15 \\
            & $^4P$+$^2F$+$^2D$+4s$^2D$  & 2.24  &  3.01 \\
($^1D$)3$d$ & $^2G$	                     & 3.14  &  0.86 \\
            & $^2F$                      &  4.24 &  1.72  \\
            & $^2D$ + $^2P$              &  5.44 & 24.08 \\
 ($^1S$)3$d$ & $^2D$                      &  6.24 &  5.16  \\
 ($^1D$)$3d$ & $^2S$                      &  6.64 &  5.16  \\
\hline
3$p^4$($^3P$)4$d$ & $^4D$                    &  6.64 & 0.94  \\
                  & $^4F$+$^4P$+$^2F$        &  6.94 & 8.93  \\
                  & $^2P$+$^2D$	             &  7.64 & 3.29  \\
($^1D$)$4d$       & $^2G$+$^2P$+$^2D$+$^2F$  &  8.64 & 24.91 \\ 
                  & $^2S$                    &  9.34 & 1.88  \\ 
($^1S$)4$d$       & $^2D$                    & 11.04 & 7.05  \\ 
\hline
3$p^4$ ($^3P$)5$d$ & $^4D$+$^4F$+$^4P$+$^2F$       &  8.84 & 2.30 \\
                   & $^2D$+$^2P$                   &  9.34 & 2.20 \\
($^1D$)5$d$        & $^2G$+$^2D$+$^2F$+$^2P$+$^2S$ & 10.64 & 4.50 \\
($^1S$)5$d$        & $^2D$                         & 12.94 & 1.00 \\
\hline
\end{tabular}
\end{table}

The initial states of ICD are the final states of the resonant 
Auger decay following the $2p_{3/2} \to 3d$ excitation in one of 
the argon atoms (see Fig. 1). These resonant Auger final states 
are well known (see, for example ~\cite{Meyer91}).    
The candidates of the ICD initial states are summarized in Table I.  
Some of these ICD initial states are the satellite 
states from which ICD transitions were observed~\cite{Lablanquie07}.
The final states of the ICD are the lowest dicationic states of Ar$_2$
that dissociate to Ar$^+(3p^{-1}~{}^2P_{1/2,3/2})$ 
and Ar$^+(3p^{-1}~{}^2P_{1/2,3/2})$, whose weighted average energy is 31.64 eV
relative to the two neutral atoms in the ground state.  
The energy difference between the energies of the ICD initial state 
and the sum energy of the two ionic fragments 
Ar$^+(3p^{-1}~{}^2P_{1/2,3/2})$ (i.e., 31.64 eV) are given in Table I. 
This amount of energy is shared among the ICD electron and two 
fragment ions. The distribution for the energy sum of the electron kinetic energy
and the KER is illustrated in Fig.~3(a).
From the resonant Auger final states given in Table I, 
ICD and radiative decay may take place. 
The radiative decay is expected to be much slower than ICD. 
Assuming that the these resonant Auger final states decay via only ICD 
as long as ICD is energetically open, the intensities of the individual 
ICD transitions are expected to be proportional to the populations 
of the ICD initial states.  
The populations can be estimated by the intensity ratios
of the resonant Auger transitions that can be found 
in literature~\cite{Meyer91} as given in Table I.  
The positions of the vertical lines in Fig. 3(b)
correspond to the expected energy sums for the ICD transitions
and their heights correspond to the relative intensities 
of the corresponding resonant Auger transitions (see Table I). 
The intensity distribution for the ICD transitions thus estimated agree 
well with the present observation, as seen in Fig. 3. 
This agreement confirms that the resonant Auger final states are 
subject to ICD as long as ICD is energetically open and that 
other competing processes such as radiative decay 
is much slower than ICD. 

The peak at 5.4 eV is assigned to the transition 
Ar$^+(3p^{-2}(^1D)3d~^2D)$-Ar $\to$
Ar$^+(3p^{-1})$-Ar$^+(3p^{-1})$, 
whereas the peak at 8.6 eV is assigned to the transition 
Ar$^+(3p^{-2}(^1D)4d~^2D)$-Ar $\to$
Ar$^+(3p^{-1})$-Ar$^+(3p^{-1})$.  
In Fig. 2(a), these two ICD transitions are indicated by 
the straight lines A and B of slope $-1$ with constant values of 
5.4 and 8.6 eV for the sum of electron kinetic energy and KER.  
As noted above, the ICD transition Ar$^+(3p^{-2}(^1D)3d~^2D)$-Ar $\to$
Ar$^+(3p^{-1})$-Ar$^+(3p^{-1})$ [A in Fig. 2(a)] takes place at the 
equilibrium internuclear distance $\sim 3.8$~{\AA} of the neutral argon dimer. 
This suggests that both the resonant Auger decay and the ICD are much 
faster than the vibrational motion of the argon dimer.  
On the other hand, the ICD transition Ar$^+(3p^{-2}(^1D)4d~^2D)$-Ar $\to$
Ar$^+(3p^{-1})$-Ar$^+(3p^{-1})$ [B in Fig. 2(a)] takes place at the 
internuclear distance $\sim 2.8$~{\AA} that is significantly shorter
than the equilibrium distance. This clearly illustrates that the ICD from 
Ar$^+(3p^{-2}(^1D)4d~^2D)$-Ar is significantly slower than that from 
Ar$^+(3p^{-2}(^1D)3d~^2D)$-Ar and takes place after bond-shrinking.  
It is worth noting that the ICD transition from Ar$^+(3p^{-2}(^1D)4d~^2D)$-Ar 
does not occur at nuclear distances smaller than $\sim~2.8$~\AA, 
though ICD is in principle energetically allowed there. 
Therefore, the internuclear distance 2.8~{\AA} may correspond to 
the inner turning point of the classical vibrational motion for 
Ar$^+(3p^{-2}(^1D)4d~^2D)$-Ar.  A recent theoretical study 
estimated that the turning point is $\sim 2.6$~{\AA}~\cite{theory}, 
supporting our hypothesis.  The ICD rate increases 
rapidly with the decrease in the nuclear distance. 
Thus, if the ICD is slower than the nuclear motion, ICD takes place 
mostly at this inner turning point where the ICD rate takes the maximum. 
A classical vibrational period of the argon dimer in the ground state 
is $\sim 1$ ps.  Those of ICD initial states may be shorter. 
The travel time for the vibrational wavepacket to reach 
the inner turning point is therefore $\sim 0.5$ ps at most. 
Thus, the lifetime of Ar$^+(3p^{-2}(^1D)4d~^2D)$-Ar is expected 
to be longer than 0.5 ps while that of Ar$^+(3p^{-2}(^1D)3d~^2D)$-Ar 
should be much shorter than 0.5 ps. These different lifetimes 
were confirmed by the recent theoretical study that 
estimated the lifetimes $14-46$ fs for Ar$^+(3p^{-2}(^1D)3d~^2D)$-Ar 
and $0.5-1.4$ ps for Ar$^+(3p^{-2}(^1D)4d~^2D)$-Ar~\cite{theory}.

Relevance of ICD and radiation damage was noted several times 
in literature~\cite{Morishita06,Jahnke10,Mucke10,Vendrell10,Stoychev11,Kim11}
due to the following reason. 
Genotoxic damage by high-energy ionizing radiation, including 
the breaking of DNA strands in living cells, is not caused by 
direct ionization but is induced by the secondary electrons 
produced by the primary ionizing radiation.  Bouda{\"{\i}}ffa {\em et
al.}~\cite{Boudaiffa00} found that low-energy (1 to 20~eV)
electrons can break DNA strands. Hanel {\em et
al.}~\cite{Hanel03} demonstrated that the uracil molecule, one of
the base units of RNA, is efficiently fragmented by electrons with
energies $\le 1~$eV, i.e., below the threshold for electronic
excitations.  The ICD induced by high-energy ionizing radiation
may undergo in biological molecules in the biological environment.  
Kinetic energy of the ICD electron is much lower than 
that of the first-step Auger electron. 
Thus the low-energy ICD electrons may cause radiation damage.    
In turn, radiation therapy requires the radiation damage at a specific site. 
Core excitation with a monochromatic radiation 
may be specific to the site and thus the ICD following 
the resonant Auger decay produces the low-energy electron 
that may cause damage to a specific site. 

In conclusion, we demonstrated that the most of the resonant 
Auger final states decay via ICD, using Ar$_2$ as a prototype sample
and tuning the excitation photon energy to  the $2p \to 3d$ resonance.   
Since the resonant core excitation is intrinsically site 
specific, the present study implicates a new way to produce 
low energy electrons at high efficiency, causing local radiation 
damage at a specific site.  
We found also that the lifetime of Ar$^+(3p^{-2}(^1D)4d~^2D)$-Ar 
is longer than the lifetime of the Ar$^+(3p^{-2}(^1D)3d~^2D)$-Ar.

We are grateful to K. Gokhberg, A.I. Kuleff, P. Demekhin, and L.S.
Cederbaum  for stimulating discussion, K. Gokhberg for providing us 
with theoretical results prior to publication, U. Hergenhahn for critical 
reading the manuscript, and to K. Nagaya and M. Yao for their 
instruction of the sample beam preparation.
The experiments were performed at SPring-8 with the approval of JASRI.
The work was supported by Grant-in-Aid (21244062) from JSPS 
and by the Management Expenses Grants for National Universities
Corporations from MEXT. 

Note added: The CID after resonant Auger decay in the N$_2$ and CO dimers
has been very recently independently observed~\cite{Trinter}.

\bibliographystyle{ieeetr}

\begin{thebibliography}{10}

%
\bibitem{Cederbaum97}
L.S. Cederbaum, J. Zobeley, and F. Tarantelli, Phys. Rev. Lett.
\textbf{79}, 4778 (1997).
%
\bibitem{Averbukh}
V. Averbukh \emph{et al.}, 
J. Electr. Spectr. Relat. Phenom. \textbf{183}, 36 (2011).
%
\bibitem{Hergenhahn}
U. Hergenhahn, J. Electr. Spectr. Relat. Phenom {\bf 184}, 78 (2011). 

\bibitem{Jahnke07}
T.~Jahnke \emph{et al.},
Phys. Rev. Lett. \textbf{99}, 153401 (2007).
%
\bibitem{Lablanquie07}
P. Lablanquie \emph{et al.},
J. Chem. Phys. \textbf{127}, 154323 (2007).
%
\bibitem{Havermeier10}
T. Havermeier \emph{et al.},
Phys. Rev. Lett. \textbf{104}, 133401 (2010).

\bibitem{Sisourat10}
N. Sisourat  \emph{et al.},
Nature Physics \textbf{6}, 508 (2010).




\bibitem{Santra03}
R. Santra and L.S. Cederbaum, Phys. Rev. Lett. \textbf{90}, 153401
(2003); \textbf{94}, 199901(E)(2005).

\bibitem{Morishita06}
Y. Morishita {\em et al.},
Phys. Rev. Lett. \textbf{96},  243402 (2006).

\bibitem{Morishita08}
Y. Morishita \emph{et al.},
J. Phys. B: At. Mol. Opt. Phys. \textbf{41}, 025101 (2008).

\bibitem{Kreidi08a}
K. Kreidi \emph{et al.},
J. Phys. B: At. Mol. Opt. Phys. \textbf{41}, 101002 (2008).

\bibitem{Yamazaki08}
M. Yamazaki \emph{et al.},
Phys. Rev. Lett. \textbf{101},  043004 (2008).

\bibitem{Kreidi08b}
K. Kreidi \emph{et al.},
Phys. Rev. A \textbf{78}, 043422 (2008).

\bibitem{Kreidi09}
K. Kreidi  \emph{et al.},
Phys. Rev. Lett. \textbf{103},  033001 (2009).

%
%
\bibitem{Barth05}
S. Barth \emph{et al.},
J. Chem. Phys. \textbf{122}, 241102 (2011). 

\bibitem{Aoto06}
T. Aoto \emph{et al.},
Phys. Rev. Lett. \textbf{97}, 243401 (2006).
%
\bibitem{Gokhberg06}
K. Gokhberg, V. Averbukh, and L.C. Dederbaum, 
J. Chem. Phys. \textbf{124}, 144305 (2006). 


\bibitem{Averbukh09}
V. Averbukh and P. Kolorenc, Phys. Rev. Lett. \textbf{103}, 183001
(2009).
%
\bibitem{Ouchi11}
T. Ouchi  {\em et al.},
Phys. Rev. Lett. \textbf{107}, 053401 (2011).

\bibitem{Zobeley01}
J. Zobeley, R. Santra, and L.S. Cederbaum, J. Chem. Phys.
\textbf{115}, 5076 (2001).
%
\bibitem{Sakai11}
K. Sakai {\em et al.},
Phys. Rev. Lett. \textbf{106}, 033401 (2011).
%
\bibitem{Forstel11}
M. F\"orstel, M. Mucke, T. Arion, A.M. Bradshaw, and
U. Hergenhahn,
Phys. Rev. Lett. \textbf{106}, 033402 (2011).


%
%

\bibitem{Jahnke10}
T. Jahnke \emph{et al.},
Nature Physics \textbf{6}, 139 (2010).
%
\bibitem{Mucke10}
M. Mucke  \emph{et al.},
Nature Physics \textbf{6}, 143 (2010).
%
\bibitem{Vendrell10}
O. Vendrell, S. D. Stoychev, and L.S. Cederbaum, Chem. Phys. Chem.
\textbf{11}, 1006 (2010).
%
%
%
\bibitem{Stoychev11}
S.D. Stoychev, A.I. Kuleff, and L.S. Cederbaum, 
J. Am. Chem. Soc. \textbf{133}, 6817 (2011). 
%
\bibitem{Kim11}
H.-K. Kim \emph{et al.},
Proc. Nat. Acad. Sci. \textbf{108}, 11821 (2011). 
%
\bibitem{Gokhberg}
K. Gokhberg, P. Kolorenc, A. I. Kuleff, and  L. S. Cederbaum, 
submitted to Science.
%
%
\bibitem{Ohashi01a}
H. Ohashi \emph{et al.},
Nucl. Instrum. Methods A \textbf{467-468}, 529 (2001).

\bibitem{Ohashi01b} 
H.~Ohashi \emph{et al.},
Nucl. Instrum. Methods A \textbf{467-468}, 533 (2001).

\bibitem{Ueda03}
K. Ueda, J. Phys. B: At. Mol. Opt. Phys. \textbf{36}, R1 (2003).

\bibitem{Kato07JES}
M. Kato \emph{et al.},
J. Electr. Spectosc. Relat. Phenom. \textbf{160}, 39 (2007).


\bibitem{Ueda08}
K. Ueda \emph{et al.},
J. Electr. Spectosc. Relat. Phenom. \textbf{166-167}, 3 (2008).

\bibitem{Ullrich03}
J. Ullrich,  
R. Moshammer, A. Dorn, R. D{\"o}rner,
L.Ph.H. Schmidt, H. Schmidt-B{\"o}cking,
Rep. Prog. Phys. \textbf{66}, 1463 (2003).


%
\bibitem{Ogilvie92}
J.F. Ogilvie and F.Y.H. Wang, J. Mol. Struct. \textbf{273},  277
(1992).

\bibitem{Meyer91}
M. Meyer, E. v. Ravan, B. Sonntag, and J. E. Hansen, 
Phys. Rev. A \textbf{43}, 177
(1991).

%
%

\bibitem{Fano}
U. Fano and J. Cooper, Rev. Mod. Phys. \textbf{40}, 441 (1968).

\bibitem{Boudaiffa00}
B. Bouda{\"{\i}}ffa, P. Cloutier, D. Hunting, M. A. Huels, and L.
Sanche, Science {\bf 287}, 1658 (2000).

\bibitem{Hanel03}
G. Hanel {\em et al.},
Phys. Rev. Lett., {\bf 90}, 188104 (2003).


\bibitem{theory}
K. Gokhberg, private communication. 


\bibitem{Trinter}
F. Trinter, M.S. Sch{\"o}ffler, H.-K. Kim, F. Sturm K. Cole, N. Neumann, A. Vredenborg, J. Williams, I. Bocharova, R. Guillemin, M. Simon, A. Belkacem, A. L. Landers, Th. Weber, H.Schmidt-B{\"o}cking, R. D{\"o}rner, and T. Jahnke, submitted to Science.

\end{thebibliography}


\end{document}